\documentclass[aps,prd,twocolumn,preprintnumbers,superscriptaddress,floatfix]{revtex4}

\usepackage{multirow, graphicx,amssymb,url,mathrsfs,amsmath}
\usepackage{eucal,wrapfig,boxedminipage,setspace,subfigure}
\usepackage{amsxtra,amstext,latexsym,dsfont}

\def\IR{{\hbox{{\rm I}\kern-.2em\hbox{\rm R}}}}
\def\IB{{\hbox{{\rm I}\kern-.2em\hbox{\rm B}}}}
\def\IN{{\hbox{{\rm I}\kern-.2em\hbox{\rm N}}}}
\def\IC{\,\,{\hbox{{\rm I}\kern-.59em\hbox{\bf C}}}}
\def\IZ{{\hbox{{\rm Z}\kern-.4em\hbox{\rm Z}}}}
\def\IP{{\hbox{{\rm I}\kern-.2em\hbox{\rm P}}}}
\def\IH{{\hbox{{\rm I}\kern-.4em\hbox{\rm H}}}}
\def\ID{{\hbox{{\rm I}\kern-.2em\hbox{\rm D}}}}

\newcommand{\beq}{\begin{equation}}
\newcommand{\eeq}{\end{equation}}
\newcommand{\bea}{\begin{eqnarray}}
\newcommand{\eea}{\end{eqnarray}}

\begin{document}

\voffset 1cm

\newcommand\sect[1]{\emph{#1}---}

\title{A Holographic Description of Colour Superconductivity}

\author{Kazem Bitaghsir Fadafan}
\affiliation{ Faculty of Physics, Shahrood University of Technology,
P.O.Box 3619995161 Shahrood, Iran}

\author{Jesus Cruz Rojas}
\affiliation{ STAG Research Centre \&  Physics and Astronomy, University of
Southampton, Southampton, SO17 1BJ, UK}

\author{Nick Evans}
\affiliation{ STAG Research Centre \&  Physics and Astronomy, University of
Southampton, Southampton, SO17 1BJ, UK}

\begin{abstract}

\noindent The difficulty of describing the gauge dependent bi-quark condensate in the QCD colour superconducting phase has made it hard to construct a holographic dual of the state. To side step this problem, we argue that near the chiral restoration transition in the temperature-chemical potential plane, the strongly coupled gluons are likely completely gapped so that the colour quantum numbers of the quarks can be thought of below that gap as global indices. A standard AdS-superconductor model can then be used to analyze the fermionic gap formation.  We investigate the role of four-fermion interactions, which might be used to include the gapped QCD interactions, on the vacuum and metastable vacua of the model. It turns out to be easiest to simply relate the standard interaction of the holographic superconductor to the strength of the gapped gluons. The result is a holographic description of the QCD colour superconducting phase diagram. We take a first look at how quark mass enters and causes a transition between the colour flavour locked phase and the 2SC phase.
\end{abstract}

\maketitle

\newpage

A fermionic system at finite chemical potential is expected to develop a Fermi surface. It is known that if there is any attractive interaction between the fermions, Cooper pair condensation will occur causing superconductivity or superfluidity. This was made transparent to a particle physics audience by the renormalization group flow analysis of the papers in \cite{rgflow} (updated to a relativistic system in \cite{Evans:1998ek}). This fact leads to the natural expectation that quarks will condense in high density QCD and there has been considerable work on understanding the phase structure over a number of years (see for example the review \cite{Alford:2007xm}). Typically the preferred condensation channel is expected to break the colour gauge group so the phenomena is referred to as colour superconductivity (CSC).

At very large chemical potentials QCD is believed to become weakly coupled due to asymptotic freedom and an exact computation of the condensation pattern is possible \cite{Son:1998uk}. The more experimentally interesting case though is when the density and temperature of the quark gluon plasma are of order the strong coupling scale $\Lambda_{c}$ and here the strongly coupled nature of the problem makes precise computation tricky. Gap equation and renormalization group analysis have been done and there is a large literature on the possible phase structure as a function of $N_f$ and the quark masses \cite{Alford:2007xm}.

Over the last two decades holography has emerged as a new tool to study strongly coupled gauge theories \cite{Maldacena:1997re}. It provides the ability to rigorously compute in theories close to large $N_c$ ${\cal N}=4$ super Yang-Mills theory (including theories with quarks \cite{Karch:2002sh}) using a weakly coupled gravitational/stringy dual. The framework has been expanded phenomenologically to AdS/QCD type models of a wider space of theories \cite{Erlich:2005qh}. It has been natural throughout this period to attempt to study the CSC phase of QCD with this new tool. There are immediately a number of large obstacles though. The CSC effect is sub-leading in the large $N_c$ limit \cite{Shuster:1999tn}. The condensate depends on $N_c$ so there is no clear large $N_c$ limit. Finally the dimension 3 condensate likely breaks the gauge group yet on the gravitational side only gauge invariant operators are manifest so it is not clear how to even pose the problem (the gauge invariant square of the operator is dimension 6 but a stringy state in the dual theory). Nevertheless an instability to pair condensation of gauginos, which can form a colour singlet pair, in the presence of a chemical potential was observed, for example, early on in \cite{Evans:2001ab}. This idea was phenomenologically used to develop AdS descriptions of superconducting condensed matter systems \cite{Hartnoll:2008vx} leading to the AdS/CM field of study. Holographic studies of related instabilities in theories with scalar quarks have also been studied in \cite{Chen:2009kx}.

In this paper we want to return to the problem in QCD. The obstacles above remain grave so our approach will be to side step them. In the intermediate density phase of QCD the quark gluon plasma is strongly coupled and full of free electrically charged quarks and presumably composite, magnetically charged scalars (see \cite{Ramamurti:2018evz} for a recent discussion). These latter condense below the chiral phase transition to cause confinement (at least in the pure glue theory). Above the transition such states will still be present if not condensed. The expectation is that these fields, through loop diagrams, will generate a Debye mass of order $g \sqrt{T^2 + \mu^2}$ for both the electric and magnetic gluons (the latter are not gapped at weak coupling where there are no magnetic charges present \cite{FandM}). We will posit here that, because $g$ is large,  there can be an order of magnitude gap between the gluon mass and the chemical potential/temperature scale. We will squeeze a holographic description, in the spirit of AdS/QCD,  into this energy regime. Since the gluons are gapped we will dodge the issue of treating the SU(3) colour symmetry of the quarks as a gauge symmetry and instead impose it as just a flavour symmetry. Although the biquark condensate will further gap the gluons we presume this to be a small effect relative to the Debye screening. These assumptions will  saves us from the problems encountered holographically to date. 

 We stress that we work in a phenomenological bottom up fashion in the spirit of AdS/QCD (at $N_c=3$) or AdS/CM (where phonon interactions of electrons are described). We will use an AdS space to phenomenologically describe the conformal symmetries of the free fermions below the Debye gap scale which are then broken by the operators and sources of the theory that appear in the bulk - for example, temperature, chemical potential etc. One should of course worry that at low $N_c$ the bulk modes might become strongly coupled and stringy but the AdS/QCD philosophy is to soldier on and measure success by the output.
Our model is at heart the simplest AdS/superconductor model \cite{Hartnoll:2008vx}. We  still need to correctly describe the broken QCD interactions that generate the Cooper pair condensation! We will discuss reintroducing the interactions as four fermion terms using Witten's double trace prescription \cite{Witten:2001ua} (Recent work on developing the holography of four fermion operators can be found in \cite{Faulkner:2010gj, evanskim}). There are subtleties in this analysis including excited states of the vacuum and it turns out that an infinitely repulsive force is needed to switch off the inherent attractive channel of the base AdS/superconductivity model. We will conclude we should just choose to tune the intrinsic pairing interaction of the holographic model to represent the broken QCD interactions on the global colour degrees of freedom.  The goal of this paper is to study such broken gauge interactions in the quark gluon plasma to develop a sensible description of the CSC phase in the QCD phase diagram.

Let us quickly review the CSC condensation patterns that will interest us here \cite{Alford:2007xm}. The superconducting condensation is triggered by a chemical potential for U(1)$_B$ and the associated quark number density. In all case we are interested in the condensation of a biquark operator with quark number 2 or baryon number 2/3. We assume that at strong coupling the $\bar{3}$ colour channel remains attractive as at weak coupling whilst the 6 is repulsive so the condensation is the usual anti-symmetric $\bar{3}$ state. A spin 0 condensate is formed from an anti-symmetric combination of spins. The flavour wave function of the condensate must also therefore be anti-symmetric. First with three massless quark flavours this implies the condensate is an anti-symmetric flavour $\bar{3}$ also. We can represent this state by the matrix (we show the make up of the $\bar{3}$s of colour and flavour in terms of the constituents)
\beq \begin{array}{ccccccc} & & &\bar{R}& \bar{G} & \bar{B} & \\
&& &BG-GB& BR-RB & RG-GR & \\
\bar{u} &  sd-ds&| &\Delta_1&  &  & |\\
\bar{d} &  su-us&| && \Delta_2 &  &| \\
\bar{s} &  ud-du&| &&  & \Delta_3 &| \\
\end{array} \label{cscmatrix}\eeq
In the three flavour massless limit the expectation is that the condensate will be the diagonal as shown with all $\Delta_i$ equal - this is the colour-flavour locked state \cite{Alford:1998mk}. As the strange quark becomes massive the condensates of the top two rows ($\Delta_1, \Delta_2$) switch off and we expect to find a vev for the triplet, SU(2) flavour singlet of the bottom row ($\Delta_3$) - this is the 2SC phase of the massless two flavour case. Note all of these states carry net colour charge although we have argued the main source of gluon mass is the Debye screening rather than the Meissner induced mass. In the holographic model we will describe an AdS-scalar $\psi$ that is dual to an element $\Delta_i$ of this matrix which acquires a vev. We will seek the phase boundary where the condensate switches on in the $T - \mu$ plane. We will briefly discuss including a quark mass in our final section to display a transition between the colour-flavour locked and the 2SC phases although as we will stress the analysis is very naive and challenges remain to find a complete holographic picture.

In Section I we will review the origin of the electric and magnetic Debye gluon masses that generate a gap. In Section II we review the AdS superconductor model that we will use including fields for each of the biquark gaps we consider. In Section III we look at the role of four quark operators in the supercondutor model including the role of unstable minima of the model.  In Section IV we match the superconductor model's coupling to the QCD coupling inthe $T-\mu$ plane to predict the gap size. In sub-section IV.1 we discuss how quark mass would enter the holographic model to suppress the biquark condensates. Finally in Section V we conclude. 

\section{Electric and Magnetic Debye Masses}

Our arguments about the gapping of the gluonic degress of freedom are important to our approach so we will briefly review the ideas already in the literature in more detail. At high density QCD is believed to become weakly coupled  and one can explicitly compute in perturbation theory \cite{FandM}. Here it is then known that the electric $A_0$ gluon components acquire a Debye mass of order $g \mu$ . The magnetic $A_i$ degrees of freedom though are not fully screened but instead Landau damped.  Their self energy behaves as $\Sigma^2 \sim  g^2 \mu^2 |q_0| /|q|$ with $q$ the gluonic four momentum.  In such weakly coupled theories if colour superconductivity sets in then the charged gap is the only source of mass for the magnetic gluonic degrees of freedom. Indeed a central point of the  analysis in \cite{Son:1998uk} was to include the effects of the Landau dampng in the estimate for the gap scale.

 We argue though that at low chemical potential, which is relevant for neutron star and heavy ion collisions, the behaviour is probably rather different. In particular we expect the QCD plasma to contain magnetically charged  scalars (see \cite{Ramamurti:2018evz} for a recent discussion) because of their role in confinement below the chiral/deconfinement transition. If such (composite) states do exist then they will simply through one loop diagrams generate a Debye like mass for the magnetic $A_i$ gluonic degrees of freedom too. Now all the gluons are gapped at the scale $g \mu$ which for chemical potentials in the hundreds of MeV and for $g \sim 4 \pi$ are much higher than the superconductor gap scale which is typically estimated in the 10s of MeV.  This separation of scales motivates a description of colour supeconductivity in which the quarks exist as the sole degrees of freedom in the low energy theory below $g \mu$ interacting only by four fermion operators generated by the gluons. In such a description the colour quantum numbers of the quarks will appear as global quantum numbers (although a full description of all higher dimension operators would secretly include gauge invariance). Since holographic colour superconductor models describe the breaking of global symmetries we can now hope to apply that framework to this energy regime in high density QCD. Note that we assume that the contribution ot the gluon gap from the low scale superconducting condensate is small relative to the Debye masses generated by the plasma so that the cut off scale and gap can be considered disconnected.

\section{AdS Superconductors}

As a first start in this paper, we will just address the CSC phase which is the novel physics of interest. We will assume the chiral transition, where the $\bar{q}q$ condensation occurs, is at the scale where the QCD coupling diverges $\Lambda_c$. Thus consider $T,\mu$ scales above this energy scale only. One finds the phase diagram in Figure 5.

Let us begin by setting up a very simple AdS description of superconductivity following the start up model of \cite{Hartnoll:2008vx}. We place our description in a black hole geometry (which we will not backreact)
\beq
ds^2 = {r^2 } ( - f dt^2 + d\vec{x}^2) +  {1\over r^2 f} dr^2, \hspace{1cm}  f = 1 - {r_H^4 \over r^4}.
\eeq
Here $\vec{x}=x,y,z$ are the boundary coordinates and the radial distance is $r$ so that the boundary is located at infinity. The usual relation between temperature and the horizon position is $r_H = \pi T$ (we have set the AdS radius to 1).

The key ingredients we need are a scalar field, $\psi_i$, to represent the quark bilinear $\Delta_i$ from (10 (here a component of the quark bilinear in the $\bar{3}$ of colour) with baryon number $B= 2/3$ and dimension 3, and a gauge field associated with U(1)$_B$ whose $A_t$ component will describe the chemical potential. We use an action
\beq {\cal L} = -{1 \over 4} F^{\mu \nu} F_{\mu \nu} - | \partial \psi_i - i B A \psi_i |^2 + 3 \psi_i^2 ,\eeq
the mass is picked to be minus three in units of the AdS radius since this corresponds holographically via $M^2 = \Delta(\Delta-4)$ to a dimension 3 operator. We have neglected any order $\psi^4$ interaction terms between different $\psi_i$.

The equations of motion are
\beq \psi_i'' + \left({f' \over f} + {5 \over r} \right) \psi_i' + { B^2 \over r^4 f^2} A_t^2 \psi_i
+ {3 \over r^2 f} \psi_i=0,  \label{eom1}
\eeq
and
\beq
A_t'' + {3 \over r} A_t' - \sum_i {2 B^2 \over r^2 f} \psi_i^2 A_t = 0. \label{eom2}\eeq

As usual for regularity one requires $A_t=0$ at the horizon which implies from the first equation of motion that
\beq \psi_i' = -{3 \over 4 r_H} \psi_i. \eeq
Note that strictly at $T=0$ we can not assume this boundary condition and the model is not complete. We will use the model to work out the edge of the phase boundary at finite T and not address the T=0 state.

There is always a solution
\beq \psi_i=0, \hspace{1cm}  A_t = \mu - {\mu r_H^2 \over r^2}. \eeq

There are more complex solutions that we can find numerically by shooting out from the horizon. In the UV they take the form
\beq \psi_i = {J_c \over r} + {c \over r^3}+... \hspace{1cm} A_t = \mu + {d \over r^2}+... \label{UVform} \eeq
$c$ is interpreted as the Cooper pair condensate, ${\cal O}=\psi \psi$, $J_c$ the source for that operator (which carries both colour and flavour indices generically), $\mu$ is the chemical potential and $d$ the density. Since there are two constraints on $\psi, \psi', A_t, A_t'$ at the horizon we get a two parameter family of solutions (set in the IR by $\psi(r_H)$ and $A_t'(r_H)$ which we label by the values of $J_c$ and $\mu$, predicting $c$ and $d$).

\begin{figure}[]
 \centering
{ \includegraphics[width=7.5cm]{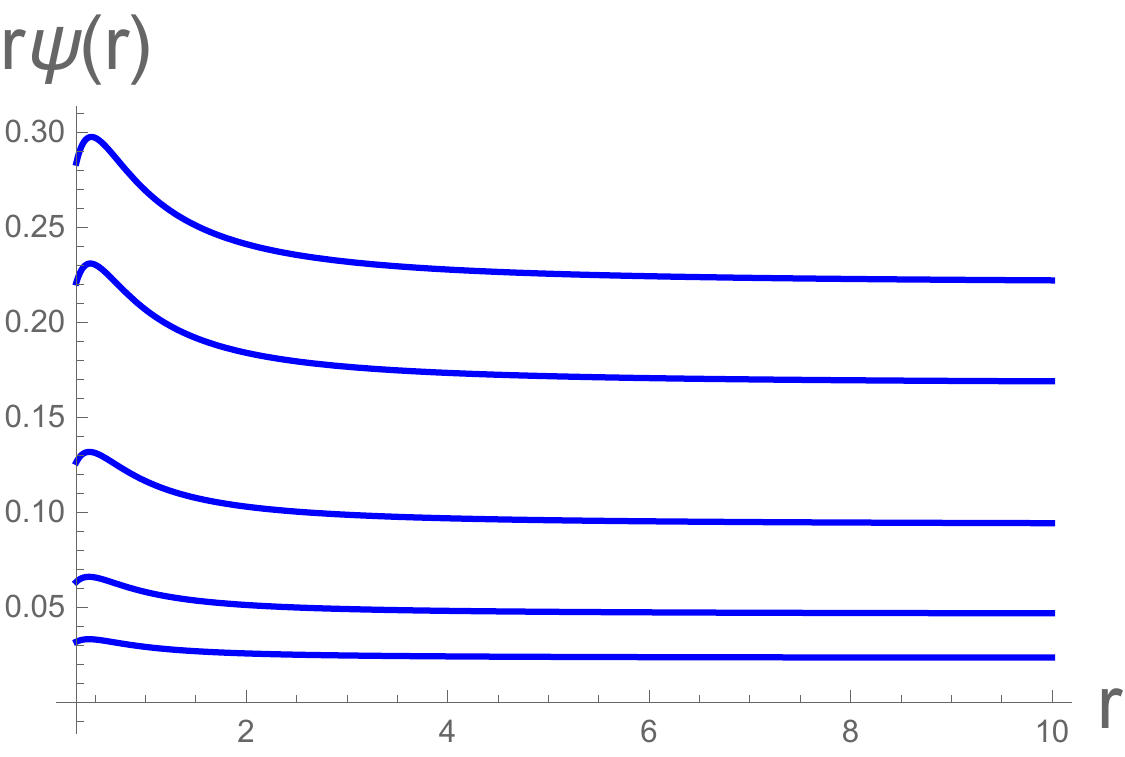}  \\ (a) \\\includegraphics[width=7.5cm]{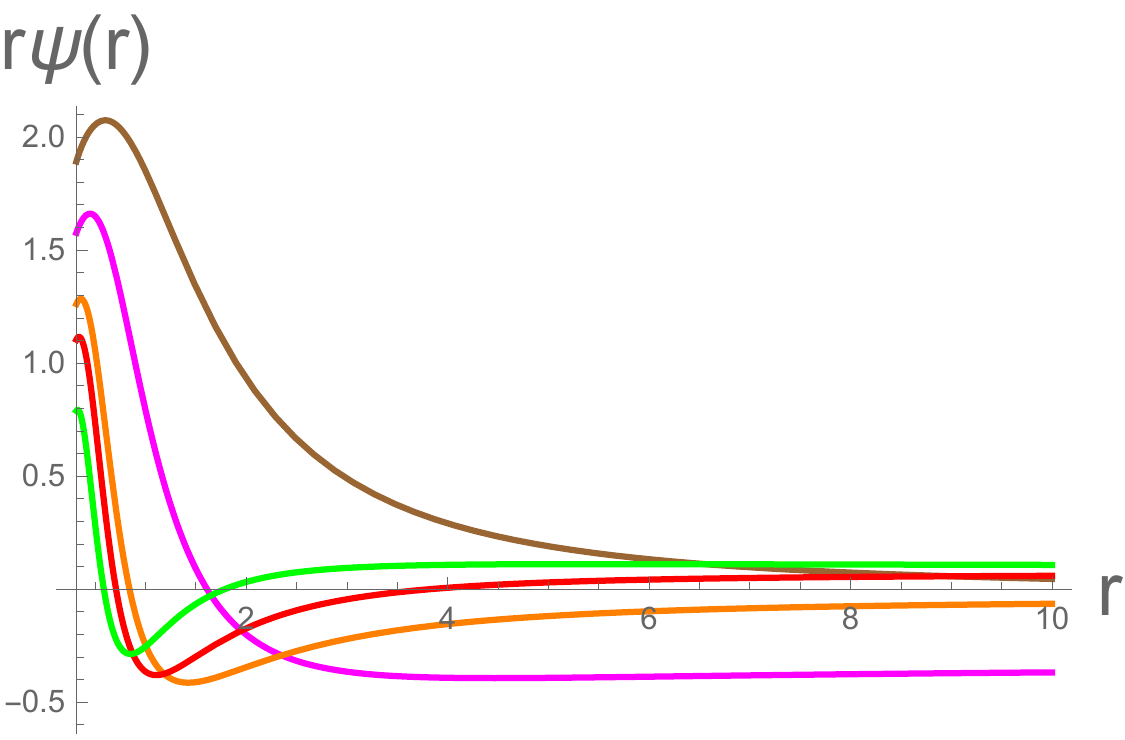} } \\ (b)
  \caption{ (a) The $\psi$ functions in the unbroken phase at $T=0.1, \mu=1.0$. (b) The $\psi$ functions in the broken phase at $T=0.1, \mu=5.0$.}
            \label{fig:1}
\end{figure}

For example, let us consider the case with a single $\psi_i$ field which might describe $\Delta_3$ in the two flavour case.  Note that if there is more than one identical $\psi_i$ then there is an effective factor of $N_i$ in the interaction term in (5). This can be removed by rescaling the $\psi_i$ by $1/ \sqrt{N_i}$ leaving the same equations to be solved. In practice this means the CFL condensates will be a factor of $\sqrt{3}$ smaller than the 2SC computations we make. Crucially though the phase boundaries remain at the same coupling values. For this reason we wil mainly study the $N_i=1$ case.

Now we can solve (4) and (5) numercially:  in Figure 1a for $T=0.1$ we plot the solutions of $\psi$ (we plot $r\psi$ which asymptotes to $J_c$ in the UV) where in each case $A_t'(r_H)$ has been adjusted to set $\mu=1.0$. In Figure 1b we show solutions for $\mu=5.0$. At low $\mu$ there is no symmetry breaking - the only solution with $J_c=0$ is that with $\psi=0$ so that $c=0$. For the higher value of $\mu$, the solution that asymptotes to $J_c=0$ is symmetry breaking  (the curve shown with the highest IR value)- this solution has a non-zero condensate $c$. 
the physics here in AdS is that the chemical potential generates an effective negative mass squared for the scalar $\psi$ and when it violates the Brietenlohner-Freedman (BF) bound of $M^2 = -4$ \cite{Breitenlohner:1982jf} an instability to $\psi$ condensation results. 

In Figure 2 we plot the value of the condensate against $\mu$ for the $J_c=0$ embeddings at fixed $T=0.1$ and show there is a second order transition. Note that the presence of this transition means the model has an intrinsic attractive interaction built into it - condensation would not occur otherwise. Below we will investigate switching off this intrinsic attraction by switching on a repulsive four fermion interaction but also move to adjusting its strength to play the role of the QCD interactions.

\begin{figure}[]
 \centering
{\includegraphics[width=7.5cm]{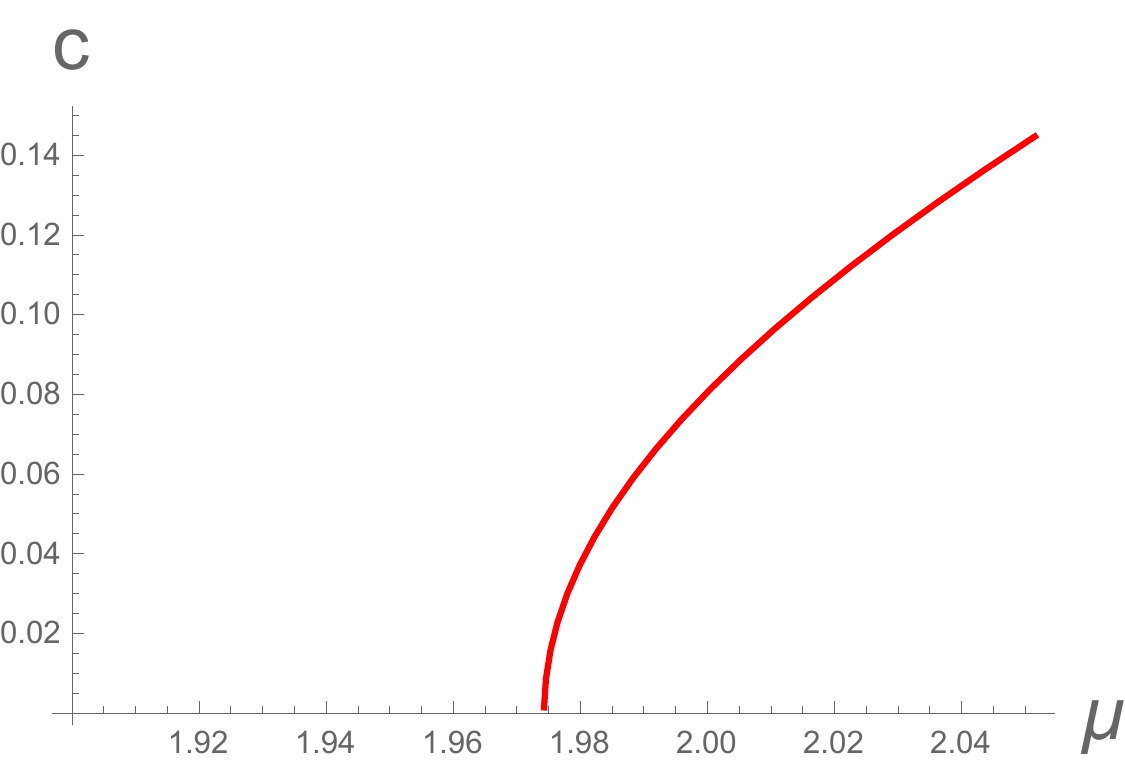}   }
  \caption{ The condensation vs $\mu$ in the broken phase at $T=0.1$. }
            \label{fig:1}
\end{figure}

The model has interesting structure beyond the basic transition. If in the broken phase we allow $\psi(r_H)$ to fall below the value that generates the $J_c=0, c \neq 0$ solution there are solutions, shown in Figure 1b, that asymptote to negative $J_c$. A minimum $J_c$ is encountered as one lowers $\psi(r_H)$ and the UV value of $J_c$ then rises again. There is a further solution with $J_c=0, c \neq0$ where the $\psi$ function dips once below the axis. This is an excited state of the vacuum where the first radially excited state of the bound states associated with $\psi$ has condensed rather than the ground state. As $\psi(r_H)$ falls further excited states can occur, with condensation of higher and higher excitation modes. We demonstrate this by plotting the solutions in the $J_c,c$ plane for $\mu=5,10$ in Figure 3 where a spiral structure is revealed. As the spiral moves between quadrants of the plane the solutions for $\psi$ change - first there are solutions for which $\psi$ is always positive, then when the solution falls below the axis in the UV we switch to negative $J_c$ and so on.  This is typical in holographic models of symmetry breaking having first been identified in the D3/D7 system with a magnetic field \cite{Filev:2007gb}. These extra vacua will play an interesting role in the discussions to come.

\begin{figure}[]
 \centering
{\includegraphics[width=7.5cm]{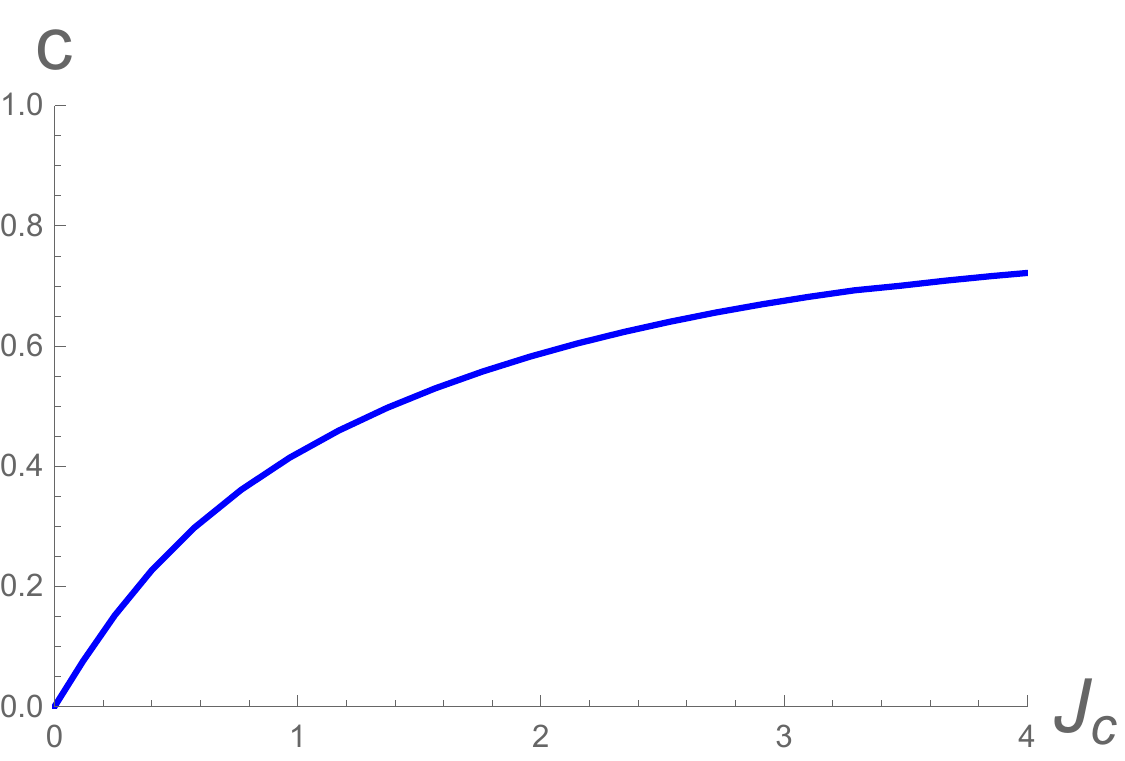} \\ (a) \\ \includegraphics[width=7.5cm]{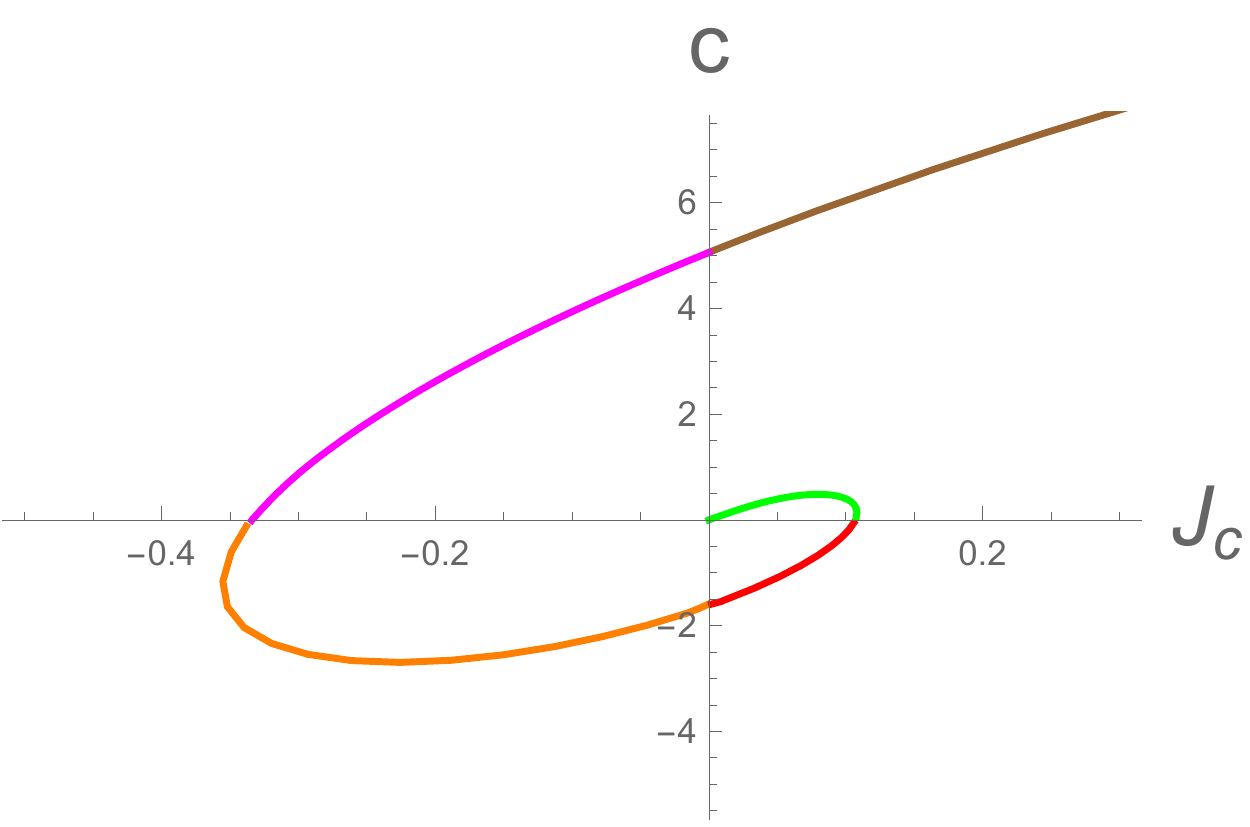} \\ (b) \\ \includegraphics[width=7.5cm]{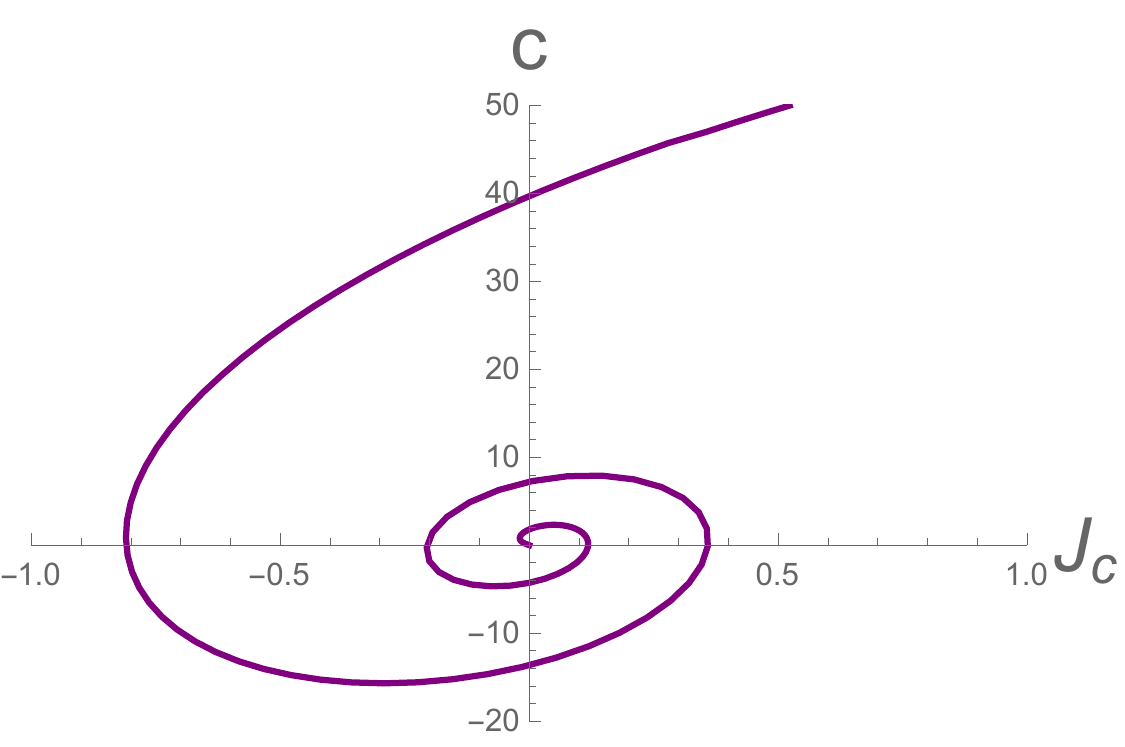} \\ (c)  }
  \caption{$c$ vs $J_c$ where $T=0.1$. (a)  Unbroken phase where $\mu=1.0$.  (b): Broken phase where $\mu=5.0$.  (c) Broken phase where $\mu=10.0$. }
            \label{fig:1}
\end{figure}

We will next turn to introducing NJL interactions into the model. 

\section{NJL Operators}

A natural step is to include the QCD interactions into our model of CSC as four fermion operators since the gluons are assumed to have acquired a large mass. Four fermion operators are an example of a ``double trace'' operator and can be incorporated using Witten's prescription \cite{Witten:2001ua}. Previous work on NJL operators in holographic superconductors can be found in \cite{Faulkner:2010gj} and recent work understanding the holographic description of the relativistic Nambu-Jona-Lasinio model is in \cite{evanskim}.

Consider the holographic description of an operator/source pair  ${\cal O}, J$ by a holographic field $\psi$ in AdS
\beq ds^2 = r^2 dx_{3+1}^2 + {dr^2 \over r^2}, \eeq
with action (here we pick $M^2=-3$ since all our operators are dimension 3)
\beq S = - \int dr {1 \over 2} \left( r^5 (\partial_r \psi)^2  - 3 r^2 \psi^2 \right). \label{psiaction}\eeq
The solutions take the form
\beq \psi = J/r + {\cal O}/r^3. \label{sol} \eeq
Evaluating the action there is a UV divergence so we must include the counter term at the UV boundary ($\Lambda$)
\beq {\cal L}_{UV} = - {1 \over 2} \Lambda^4 \psi^2|_\Lambda. \eeq
This term is crucial for the analysis below.

(Note that in the previous paper \cite{evanskim} we worked with a rescaled field $L = r \psi$. This is natural from the point of view of the D3/probe D7 system where the UV action takes precisely this form.  If one substitutes this rescaled field into the action above and integrates by part then the surface term vanishes and the action takes the form
\beq {\cal L} = - \int dr {1 \over 2} r^3 (\partial_r L)^2,  \eeq
which, since $L\sim J+ ..$, has no UV divergence and hence no counter term. The IR boundary condition $\partial_r L=0$ forces ${\cal O}=0$ which is appropriate for supersymmetric gauge theory configurations where, for example,  the quark condensate is forbidden. Here the action also vanishes with $L= {\rm constant}$ corresponding to the vacuum energy of the gauge theory vanishing.)

We now wish to include in the field theory a term of the form
\beq \Delta {\cal L} = - {g^2 \over \Lambda^2} {\cal O} {\cal O}, \label{nine} \eeq
where ${\cal O} \neq$ 0 then this term generates a source $J = {g^2 \over \Lambda^2} {\cal O}$. If we substitute this relation back into the Lagrangian term we uncover
\beq \Delta {\cal L} = - {\Lambda^2 J^2 \over g^2}.  \eeq
In analogy to this term Witten's prescription in the holographic description is to add a UV surface term evaluated at the cut off $\Lambda$
\beq \Delta {\cal L} = - {\Lambda^4 \psi^2 \over g^2},  \eeq
since $\psi \sim J/ \Lambda+..$ in the UV these match.

The simplest way to include this extra term in the analysis is by considering the result of the change to the UV boundary condition on the solutions. Varying the action gives
\begin{equation}  \delta S = 0 = -\int d r\left(\partial _r {\partial {\cal L }\over \partial \psi'}  - {\partial {\cal \psi }
\over \partial \psi} \right)  \delta \psi  + \left. {\partial {\cal L }\over \partial \psi'} \delta \psi \right|_{{UV, IR}} \,.  \end{equation}
There is also the variation of the surface counter term
\beq \delta S =  - 2 \Lambda^4 \psi \delta \psi|_{UV}. \eeq
Normally in the UV one would require the source to be fixed and $\delta \psi=0$ to satisfy the boundary condition. We do this by fixing the source $J$ to specify a particular theory.

\begin{figure}[]
 \centering
{\includegraphics[width=7.5cm]{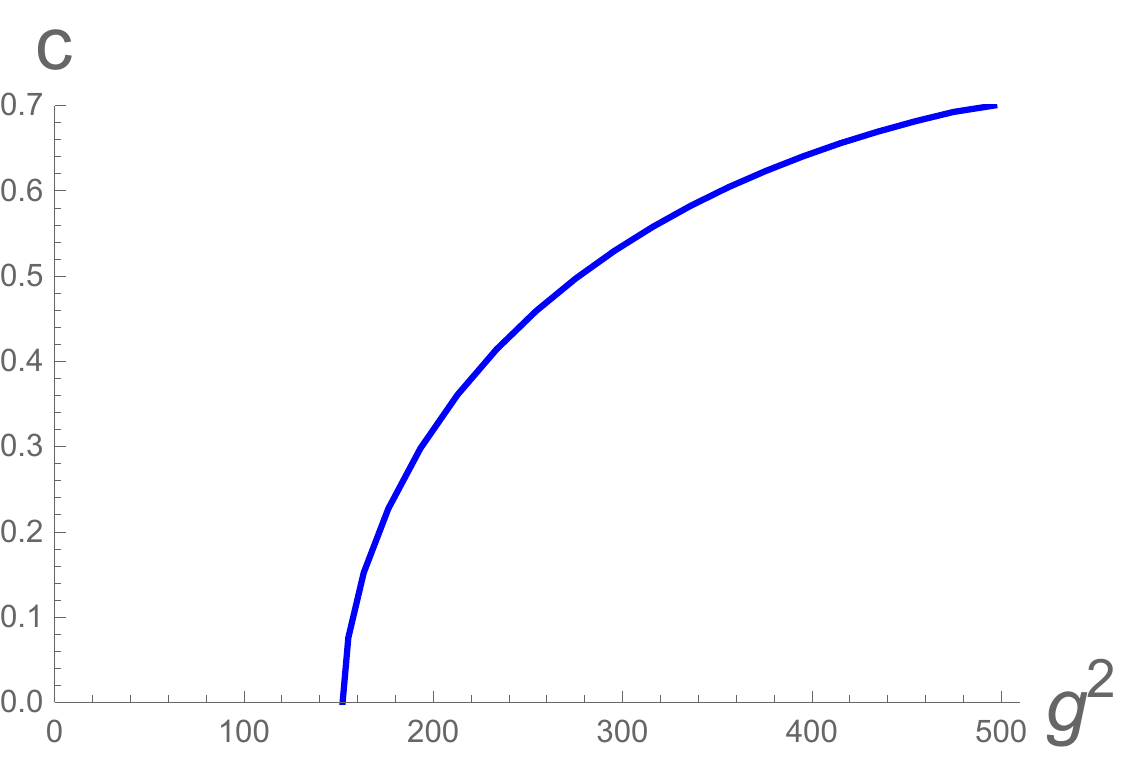} \\ (a) \\ \includegraphics[width=7.5cm]{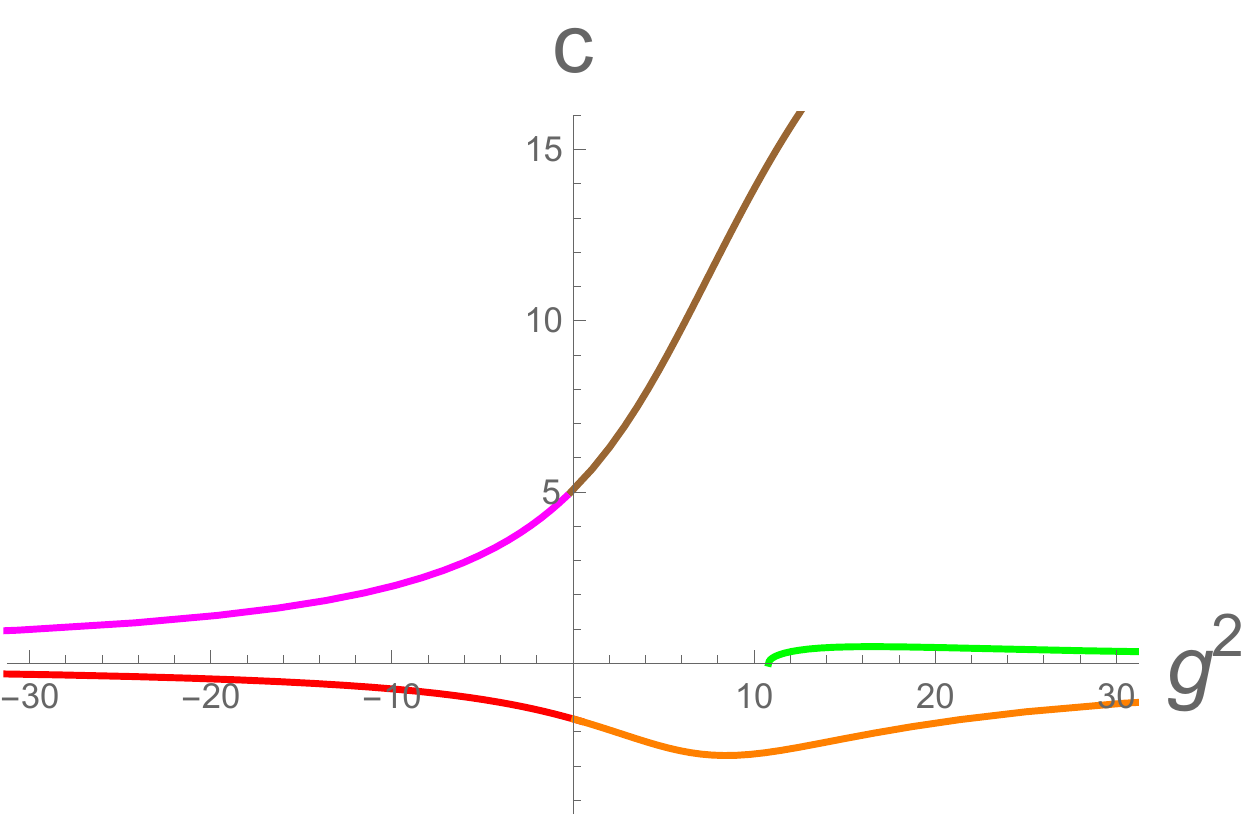} \\ (b)   }
  \caption{(a) Plot of $c$ against $g^2$ ($\Lambda=10$) in the unbroken phase for embeddings in Fig 1a. (T=0.1, $\mu=0.1$) (b) Plot of $c$ against $g^2$ ($\Lambda=10$) in the broken phase (T=0.1, $\mu=5$) for solutions in Fig 1b.}
           \label{fig:1}
\end{figure}

To describe the double trace operator though we allow $\psi$ ($J$) to change at the UV boundary and instead impose the vanishing at that boundary of
\begin{equation} 0 = {\partial {\cal L} \over \partial \psi'}  +  \psi \Lambda_{UV}^4  + {2 \psi \Lambda_{UV}^4 \over g^2 }  \,,   \end{equation}
where we have included the variation of the new surface term. For our action ${\partial {\cal L} \over \partial \psi'} =  - r^5 \psi'$. Assuming (\ref{sol}) we find that we need
\begin{equation} J \simeq {g^2 \over \Lambda^2} {\cal O}. \label{condition} \end{equation}
This condition (which matches the expectation under (\ref{nine})) is simple to apply to the solutions of the (unchanged) equation of motion we already have. \\

\noindent {\bf III.1 NJL Operators in the Superconductor} 

Let us now return to the holographic superconductor model of the previous section. We can apply our analysis to the $\psi$ functions of Figure 1. We can interpret each function, including those with $J_c \neq 0$ as describing the model with zero intrinsic $J_c$ but a four fermion operator present. The four fermion operator in the presence of the condensate $c$ generates the UV source $J_c$. For example we can translate the functions of Figure 1a, where $\mu$ lies below the critical value, through Figure 3a, to a plot of $c$ against $g^2$ which we show in Figure 4a. Here we have taken $\Lambda=10$ numerically. We observe a critical value of the NJL coupling that triggers symmetry breaking at a second order transition. Note here there are no solutions where in the UV $J_c$ and $c$ have opposite signs - putting in a repulsive four fermion term (negative $g^2$) produces no solutions other than $J_c=0, c=0$ as one might expect.

Similarly we can translate the functions of Figure 1b through Figure 3b to the plot in Figure 4b which again shows $c$ vs $g^2$ but here at $g^2=0$ there is already symmetry breaking.  There are two interesting additional features here. Firstly there are solutions at negative, repulsive, $g^2$. This is not surprising because at $g^2=0$ there is symmetry breaking - switching on a repulsive four fermion term would be expected to reduce the condensation, and it does. The surprising feature is that the condensation does not switch off completely except at infinite repulsive interaction strength (there are solutions with zero $c$ but non-zero $J_c$ that generate infinite $g^2$ values). The intrinsic attractive interaction in the AdS/superconductor model is presumably more subtle in structure than the NJL operator which is only switching parts of the interaction off. Remember in superconductor theory any attractive interaction will result in condensation. The remaining structure in the $c-g^2$ plane is the translation of the spiral in the $c-J_c$ plane seen previously.

Our initial intention to describe QCD had been to take the basic holographic superconductor model and introduce a critically tuned repulsive NJL operator to switch off condensation at each $T,\mu$ value. On this interaction free description of the quark gluon plasma we would then add back the QCD interactions as further positive shifts in the NJL coupling strengths. We have now shown that this is not achievable because the intrinsic interactions are more subtle than the NJL interaction so an infinitely repulsive interaction is needed to switch off the base condensation. However, this approach was essentially to uninvent the wheel and then reinvent it! An equally sensible approach is to simply modify the strength of the interaction between $\psi$ and $A_t$ to reflect the QCD interaction strength. Our assumption is still that the gluons are massive in this strongly coupled phase so that we can describe the colour of quarks by a global symmetry but now the interactions will be introduced through the action
\beq {\cal L} = -{1 \over 4} F^{\mu \nu} F_{\mu \nu} - | \partial \psi - i G B A \psi |^2 + 3/L^2 \psi^2. \label{nowwithG}\eeq
We interpret the $A \psi$ interaction term as the holographic models knowledge of the broken gauge interactions. Note the inclusion of the new coupling $G$ which we will shortly relate to the QCD running coupling.

First though we can find the phase boundary for the superconducting phase as a function of $G$. For each $T$ and $G$ we make plots as in Figure 2 and then plot $\mu_c(T)$ in the plane. This is shown in Figure 5. Note that given the solutions for $G=1$ one can move to another $G$ by scaling $\psi \rightarrow G \psi$ and $A_t \rightarrow G A_t$ in (\ref{eom1}), (\ref{eom2}) so the critical $\mu$ just scales with $G$.

\begin{figure}[]
 \centering
{\includegraphics[width=7.5cm]{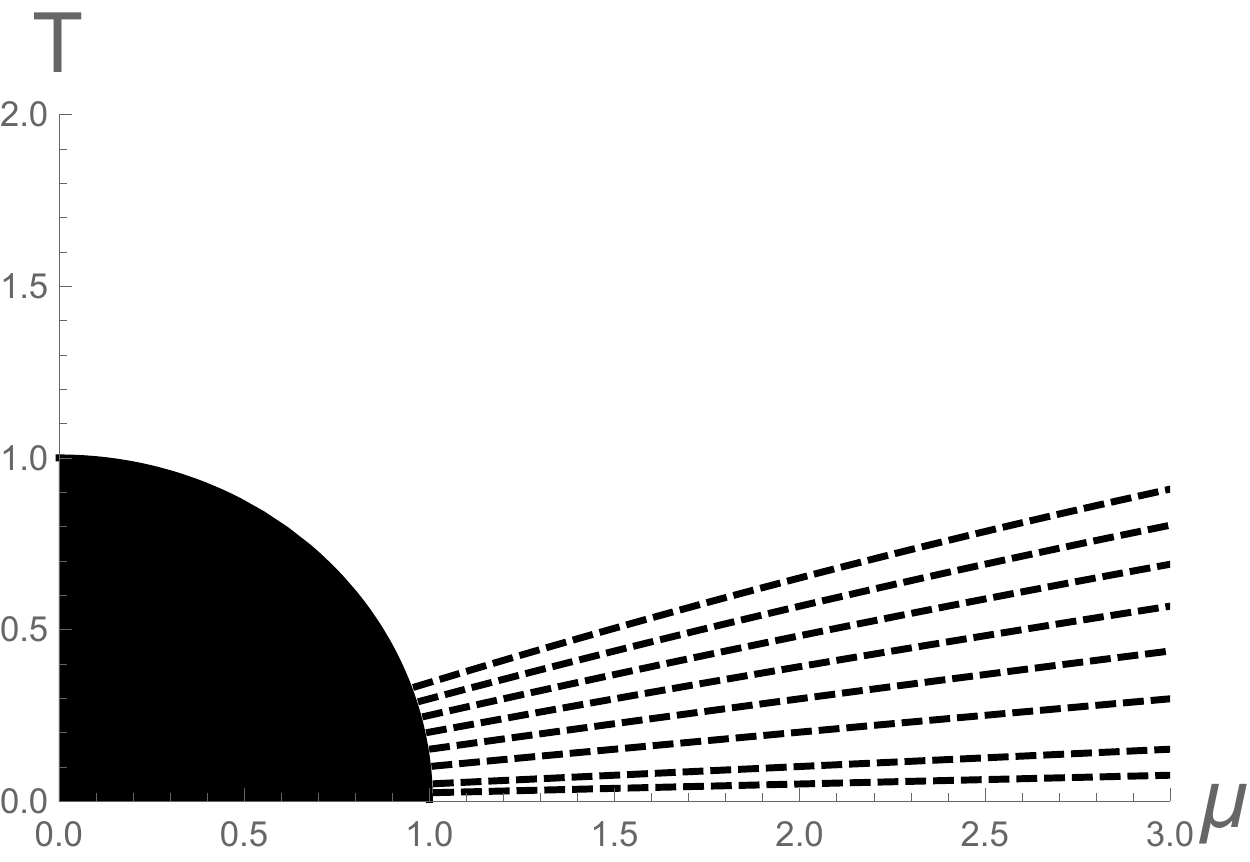}   }
  \caption{Plot of the superconducting phase boundary at different $G= 0.5,1,2,3,4,5,6,7$ from bottom to top in the T - $\mu$ plane. The black region is expected to be the chirally symmetric phase below a scale of $\mu^2+T^2 = 1$. }
            \label{fig:next}
\end{figure}

\section{The QCD Phase Diagram}

Let us now attempt to describe the colour superconducting phase of QCD using these tools. We will assume that the chiral phase transition occurs at $T^2 + \mu^2 = \Lambda_c^2$ and numerically set $\Lambda_c=1$ with a UV cut off on the holographic model of $\Lambda=10 \Lambda_c$ where we read off $c, J_c$.  We will assume a phase with a $\bar{q}q$ condensate lives below $\Lambda_c$.

In the quark gluon plasma phase we will use the action of (\ref{nowwithG}) but we must set the value of $G$ at the cut off scale to a sensible ansatz in QCD. A natural choice based on the one loop running is
\beq G^2 = {\kappa \over b \ln (T^2 + \mu^2)/\Lambda_c^2}, \hspace{1cm} b =  11 N_c/3 -  2 N_f/3, \label{qcdG}\eeq
which blows up at $\Lambda_c$. We need to fix $\kappa$ so it is appropriate for the strength of attraction that generates the $\bar{3}$ of colour condensate.

Perturbatively the strength of tree level t-channel one gluon exchange interaction for the four different colour channels for $\bar{q} q$ and $qq$ is
\beq 1_{\bar{q}q} : 8_{\bar{q}q} : 6_{qq} : \bar{3}_{qq} = -{8 \over 3} : {1\over 3} : {1 \over 6} : -{1 \over 3} \eeq
The attraction might be as little as $1/8$ the attraction for the chiral condensate. Of course at strong coupling the relative strength of these interactions is not known.
\begin{figure}[]
	\centering
	{\includegraphics[width=7.5cm]{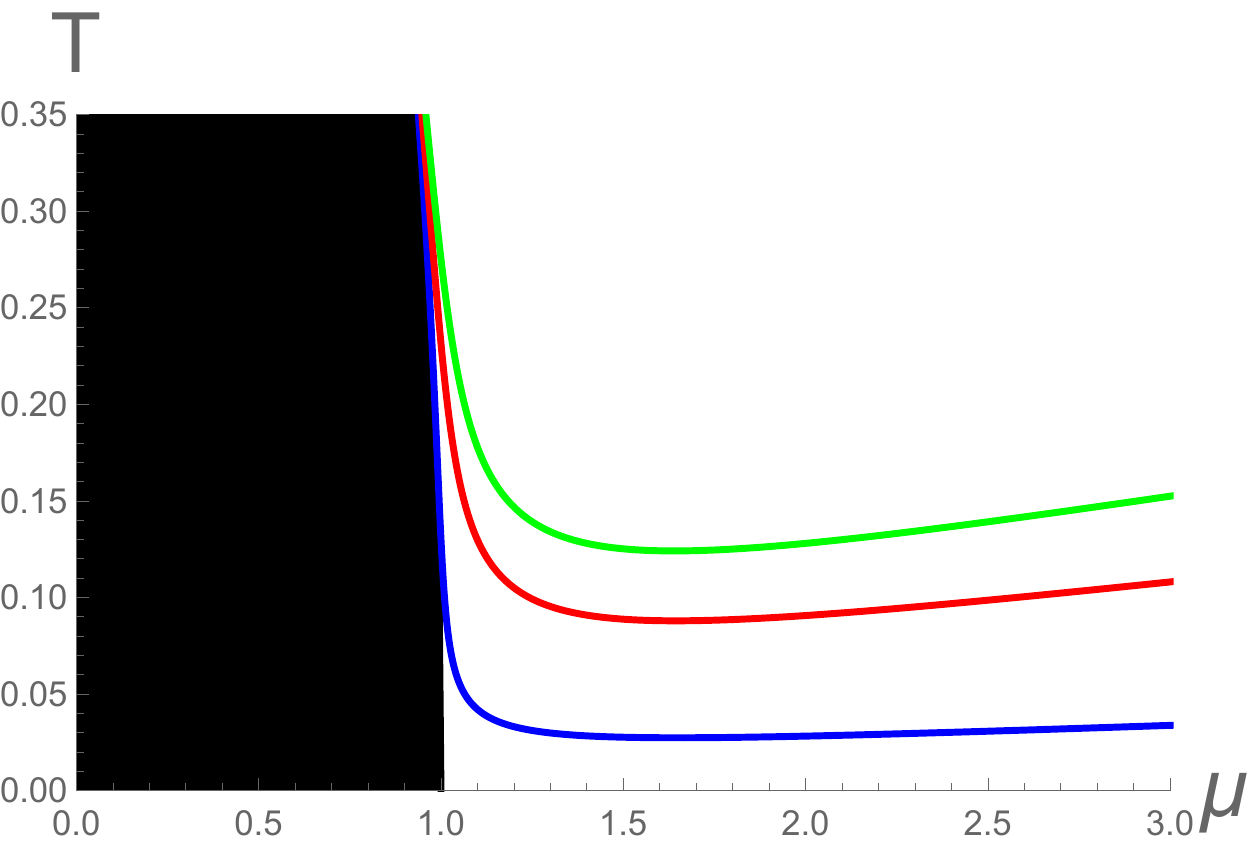}   }
	\caption{QCD phase diagram: the blacked out area is below $\Lambda_c$ where chiral symmetry breaking is expected. The remaining phase edges shows where the CFL phase is present for the choices of $\kappa=1, 10, 20$ from bottom to top......... }
	\label{fig:more}
\end{figure}
The intrinsic interaction between the fields $\psi$ and $A_t$ in the holographic model should be controlled by the strength of the QCD interactions that presumably lie between $\kappa=1 - (4 \pi)^2$. Given the 1/8th suppression we will study the range of $\kappa$ between 1 and 20 to estimate the area of the phase diagram where superconductivity is likely.

It is now simple to construct the phase diagram from the analysis of Figure \ref{fig:next}. We overlay  circles in the $T,\mu$ plane for each value of $G$ from (\ref{qcdG}) taking $N_f=3$ and identify the points where they cross the same $G$ value transition curve. We find the phase diagram in Figure \ref{fig:more}.
Very close to $\Lambda_c$ the coupling gets very strong and the superconducting phase then hugs the phase boundary up to high values of T. Most likely the chiral phase will extend a little above $\Lambda_c$ though and this feature will be greatly reduced. Typically we see the superconducting phase is predicted to exist below T of 0.15 $\Lambda_c$ (for $\kappa \simeq 10$), which we might estimate as 20 MeV or so if $\Lambda_c \simeq 175$ MeV, the expected temperature of the chiral transition. This value might rise sharply just before the chiral transition. Usual estimates place the gap in the 10-100 MeV range \cite{Alford:2007xm} so this seems a sensible model. 


\noindent {\bf IV.1 Quark Mass}  

\begin{figure}[]
 \centering
{\includegraphics[width=7.5cm]{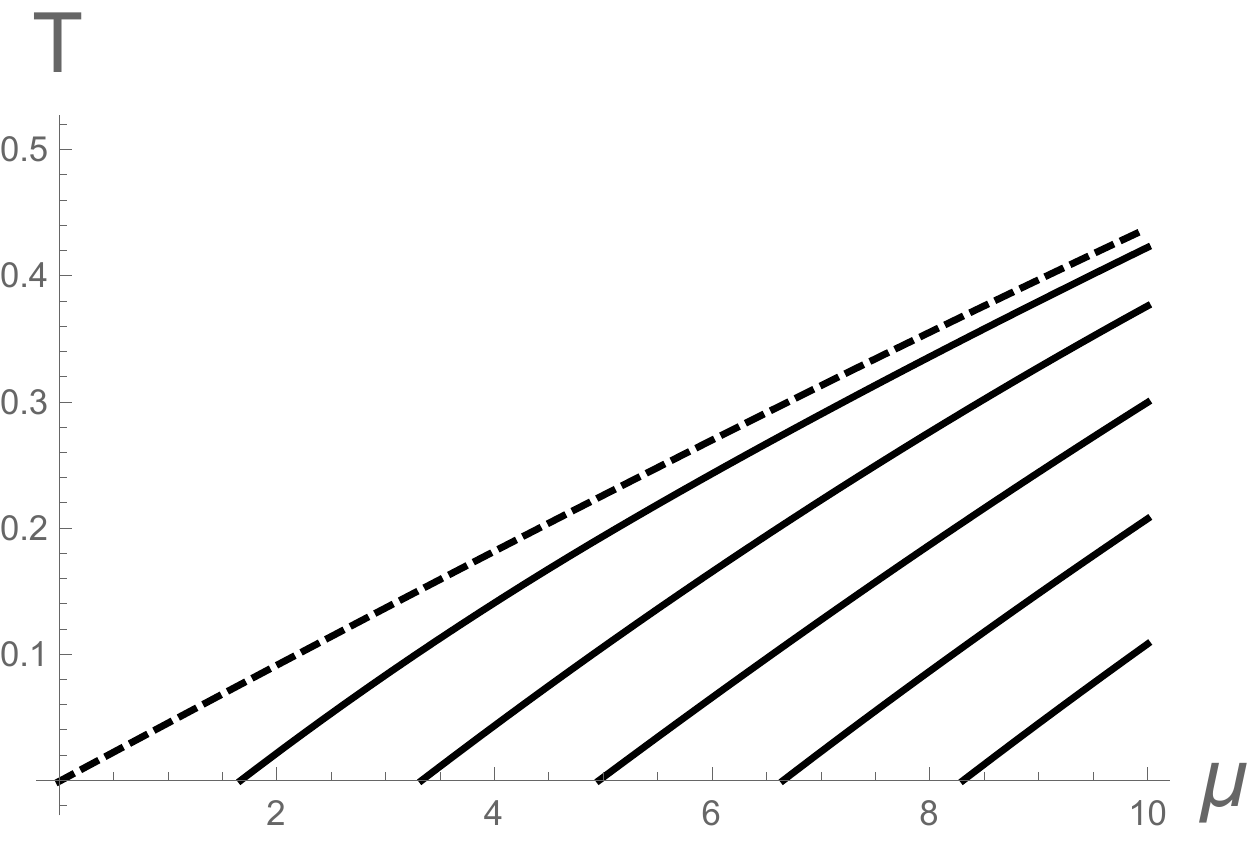}   }
  \caption{Phase diagram for the model at fixed $G = 0.9$ and for quark mass $m= 0,1.3,2,3,4,5$ from left to right.}
            \label{fig:Gmass}
\end{figure}

Our phase diagram so far has been plotted for the massless theory and the expected condensation is of the colour flavour locked form for $N_f=3$ and 2SC for $N_f=2$. One might expect there to be a transition as the strange quark mass grows from CFL to a two flavour 2SC phase at lower $\mu$. The presence of the mass leads to a lower value of the Fermi momentum which will reduce the condensate but also relative differences in the Fermi surface levels for different quark flavours is expected to frustrate the formation of the colour flavour locked condensate. 

Holographically modelling this transition is not straightfoward. Each component of (\ref{cscmatrix}) is of mixed flavour and should see the different chemical potenitals and masses associated with each of the two constituents. 
Presumably this should be described by a  non-abelian Dirac Born Infeld type action (assuming one could neglect the stringy nature of the states stretched between different flavour branes). Here though we will try something very naive to show a mechanism by which mass could switch off the condensation. 

The quark mass, $m$ should be described by a new holographic field $\chi$ with asymptotic behaviour $\chi = m/r+...$ (the solution for a scalar of mass -3 in pure AdS) and IR dynamics that should be connected to the formation of the chiral condensate which is the sub-leading operator part of the solution. One would need a full model of the chiral transition to write down a potential for the $\chi$ scalar so to avoid getting bogged down in that dynamics we will just set $\chi= m/r$ and look at its effects on the $\psi$ Cooper pair formation (of course really one should solve linked equations but our simplistic approach will show how the mass could suppress the Cooper pair  condensation). We imagine a simple Lagrangian coupling of the form $|\chi|^2|\psi|^2$ so that the equation of motion for $\psi$ becomes
\beq \psi'' + \left({f' \over f} + {5 \over r} \right) \psi' + { G^2 B^2 \over r^4 f^2} A_t^2 \psi
+ {1  \over r^2 f} \left(3 - {m^2 \over r^2}\right) \psi =0.
\eeq
Clearly the $m^2$ term acts to oppose the instability induced by $\mu$, which is the main mechanism we wish to flag here. 
\begin{figure}[]
	\centering
	{\includegraphics[width=7.5cm]{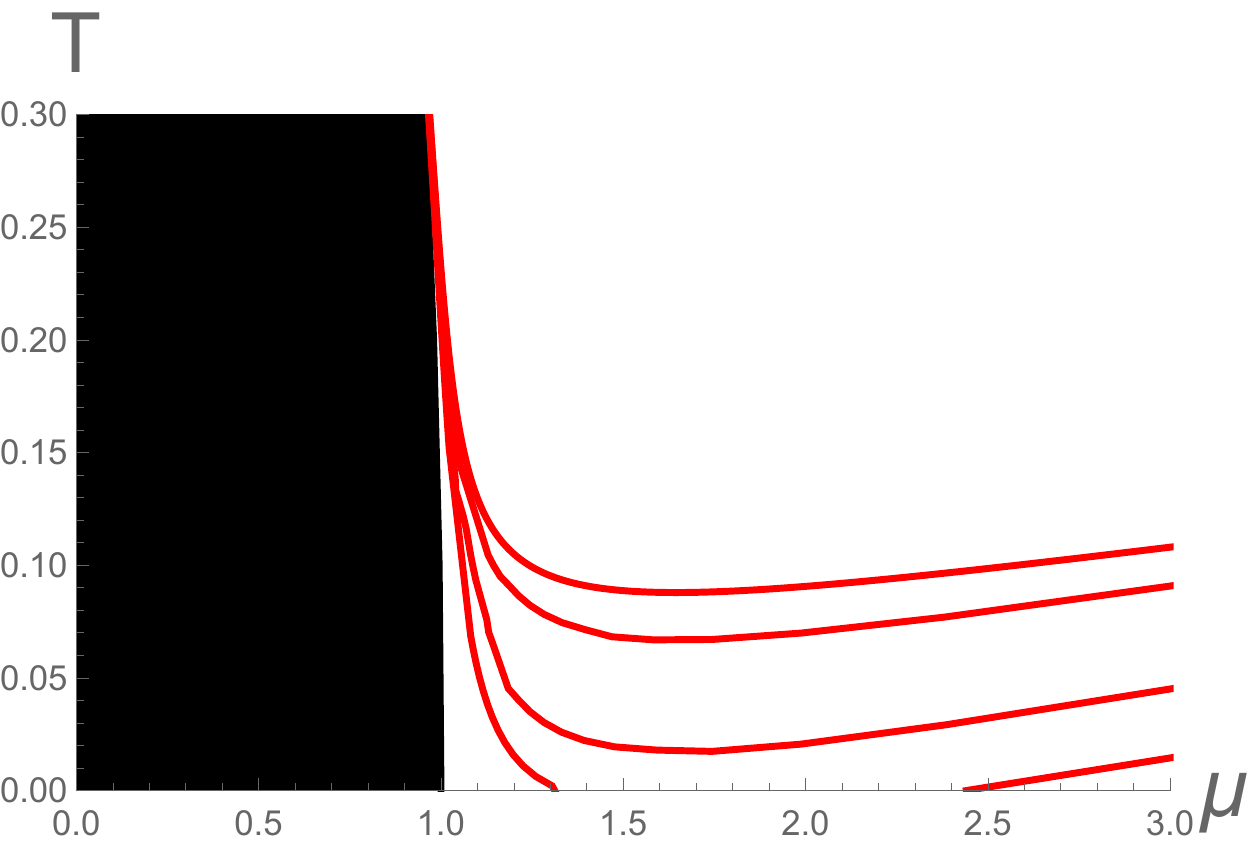}   }
	\caption{QCD phase diagram with quark mass $m$: in the blacked out area  chiral symmetry breaking is expected. The remaining phase edges shows where the CFL phase is present for the choices of $\kappa=10$ and $m=0, 0.5, 1.0, 1.3$ from top to bottom.  }
	\label{fig:mass}
\end{figure}

We first plot the phase boundary for $G=0.9$ at different values of $m$ in Figure 7. As the quark mass rises the boundary line tilts in the plane until for masses of order the chemical potential the phase is excluded at low $\mu$. The positive contribution to the scalar $\psi$'s mass squared is greater than the BF bound violating negative contribution from $A_t$. The mass therefore discourages the condensation.

We plot the phase structure of the theory with the running coupling (\ref{qcdG}) for $\kappa=10$ and $m=0.,0.5,1.0,1.3$ in Figure 8. For small quark masses the phase boundary simply moves to lower values of $T$ at a  given $\mu$. If the mass becomes larger though then for a range of $\mu$ there is no condensation present. At large $\mu$ the mass is overwhelmed and the condensation returns. Note at $\mu \simeq 1$ where the coupling becomes arbitrarily strong the phase briefly returns however large $m$ is, but this region is likely inside the chirally broken phase since the quark anti-quark attraction is also getting very strong. 

In the case of QCD one can interpret the above description as that for $\Delta_1$ and $\Delta_2$ with the interaction with $\chi$ describing the interplay with the strange quark mass.
The phase boundaries in Figure 8 for different $m$ represent our estimate of where the CFL phase ($\Delta_1$ and $\Delta_2$ ) will switch off, although it is only a naive estimate since we have not include the effect of different Fermi surface levels. The 2SC phase would be expected to exist between the edge of the CFL phase and boundary for $m=0$ since $\Delta_1$ is oblivious to the strange mass.  If we take $\kappa=10$ and assume $\Lambda_c=10$MeV then the physical strange mass corresponds roughly to the $m=0.5$ curve and the CFL phase exist down to the chiral boundary although with a transition to the 2SC phase at higher T. For lower $\kappa$ values the  CFL phase might cease completely at lower $\mu$.

This discussion has been very naive, although it reveals a mechanism by which the CFL phase will be shut down by the strange quark mass, and we leave to future work including the correct dynamics for $\chi$, including the chiral transition, as well as incorporating the  non-abelian nature of the discussion.

\section{Discussion}

The goal of this paper has been to push through the block in thinking as to how to describe the colour superconducting phase of QCD holographically. For a long while the colour charged nature of the bi-quark condensate has stopped progress. We have attempted to side step this issue by arguing that at strong coupling and intermediate temperatures and chemical potentials the gluons are likely gapped by the plasma. If we treat the colour symmetry of the quarks as a global index then holographic models can progress. Indeed here we have demonstrated that by recycling the simplest holographic superconductor model adjusted to this setting. The key question was how to then include the QCD interactions. We investigated NJL operators as one possibility and we have included the discussion here because there is an interesting story connected to the spiral structure in the operator-source plane in the superconductor model reflecting excited states of the vacuum (note that this structure is also present in the AdS$_4$ superconductor but we are not aware of any discussion of it in the literature). This leads to the conclusion that only an infinitely repulsive four fermion operator suffices to switch off the intrinsic attractive interaction of the  holographic superconductor model. That attractive interaction is presumably more complex in structure than the four fermion operator and any residual attraction would lead to superconductivity. In fact we moved to simply adjusting the strength of that intrinsic interaction to reflect the QCD couplings value as a function of $\mu,T$. As a result we can plot the phase diagram of the superconducting phase - see Figure 6. The transition temperature lies near 20MeV or so which matches the usually quoted range of 10-100MeV. 

The model we have used is somewhat like an NJL model of colour superconductivity but the holographic setting would allow one to easily compute equations of state and transport properties of the phases. We hope to investigate these and the consequences for neutron star structure and collisions in the future. There is also plenty of scope to make a more sophisticated model of the phase structure including back reaction on the metric, describing the chiral transition of QCD and the interplay between the quark mass and the condensation pattern. We made a first attempt at understanding that mass dependence by a very simple model of an interaction between a quark mass and the Cooper pair which revealed a transition between a colour flavour locked phase and a 2SC phase - shown in Figure \ref{fig:mass}. Again this matches the form of the usually expected phase structure.
\\ \\ \\

\noindent {\bf Acknowledgements:} NE's work was supported by the
STFC consolidated grant ST/P000711/1 and JCR's by Mexico's National Council of Science and Technology (CONACyT) grant 439332. KBF thanks B. Robinson for useful discussions and acknowledges the University of Southampton for their hospitality during his sabbatical.


\begin{thebibliography}{ll}


\bibitem{rgflow}
  G. Benfatto and G. Gallavotti, J. Stat. Phys. 59, 541 (199
0);  Phys. Rev. C42 (1990)
9967; R. Shankar, Physica A177, 530 (1991); Rev. Mod Phys. 66
, 129 (1993); J. Polchinski, in Proceedings of the 1992 TASI, eds. J. Harvey and J. Pol
chinski (World Scientific,
Singapore 1993)

\bibitem{Evans:1998ek}
  N.~J.~Evans, S.~D.~H.~Hsu and M.~Schwetz,
  Nucl.\ Phys.\ B {\bf 551} (1999) 275
  doi:10.1016/S0550-3213(99)00175-3
  [hep-ph/9808444].
	
\bibitem{Alford:2007xm}
  M.~G.~Alford, A.~Schmitt, K.~Rajagopal and T.~Schäfer,
  Rev.\ Mod.\ Phys.\  {\bf 80} (2008) 1455
  doi:10.1103/RevModPhys.80.1455
  [arXiv:0709.4635 [hep-ph]].
	
\bibitem{Son:1998uk}
  D.~T.~Son,
  Phys.\ Rev.\ D {\bf 59} (1999) 094019
  doi:10.1103/PhysRevD.59.094019
  [hep-ph/9812287].
	

	
	
	
\bibitem{Maldacena:1997re}
  J.~M.~Maldacena,
  Adv.\ Theor.\ Math.\ Phys.\  {\bf 2} (1998) 231  [hep-th/9711200]; E.~Witten,
  Adv.\ Theor.\ Math.\ Phys.\  {\bf 2} (1998) 253  [hep-th/9802150];  S.~S.~Gubser, I.~R.~Klebanov and A.~M.~Polyakov,
  Phys.\ Lett.\ B {\bf 428} (1998) 105  [hep-th/9802109].  
	
\bibitem{Karch:2002sh}
  A.~Karch and E.~Katz,
  JHEP {\bf 0206} (2002) 043
  [arXiv:hep-th/0205236]; M.~Grana and J.~Polchinski,
  Phys.\ Rev.\  D {\bf 65} (2002) 126005
  [arXiv:hep-th/0106014]; M.~Bertolini, P.~Di Vecchia, M.~Frau, A.~Lerda and R.~Marotta,
  Nucl.\ Phys.\  B {\bf 621} (2002) 157
  [arXiv:hep-th/0107057]; M.~Kruczenski, D.~Mateos, R.~C.~Myers and D.~J.~Winters,
  JHEP {\bf 0307} (2003) 049
  [hep-th/0304032]; J.~Erdmenger, N.~Evans, I.~Kirsch, and E.~Threlfall,   {\em Eur. Phys. J.} {\bf A35} (2008)
  81--133, [arXiv:0711.4467].
	
	
\bibitem{Erlich:2005qh}
  J.~Erlich, E.~Katz, D.~T.~Son and M.~A.~Stephanov,
  Phys.\ Rev.\ Lett.\  {\bf 95} (2005) 261602
  doi:10.1103/PhysRevLett.95.261602
  [hep-ph/0501128]; L.~Da Rold and A.~Pomarol,
  Nucl.\ Phys.\ B {\bf 721} (2005) 79
  doi:10.1016/j.nuclphysb.2005.05.009
  [hep-ph/0501218].
	
	
\bibitem{Shuster:1999tn}
  D.V. Deryagin, D.Yu. Grigoriev, and V.A. Rubakov, Int. J
. Mod. Phys. A
7
, 659 (1992); E.~Shuster and D.~T.~Son,
  Nucl.\ Phys.\ B {\bf 573} (2000) 434
  doi:10.1016/S0550-3213(99)00615-X
  [hep-ph/9905448].
	
	
\bibitem{Evans:2001ab}
  N.~J.~Evans and M.~Petrini,
  JHEP {\bf 0111} (2001) 043
  doi:10.1088/1126-6708/2001/11/043
  [hep-th/0108052].
	
	\bibitem{Hartnoll:2008vx}
  S.~A.~Hartnoll, C.~P.~Herzog and G.~T.~Horowitz,
  Phys.\ Rev.\ Lett.\  {\bf 101} (2008) 031601
  doi:10.1103/PhysRevLett.101.031601
  [arXiv:0803.3295 [hep-th]].
	
	\bibitem{Chen:2009kx}
	R.~Apreda, J.~Erdmenger, N.~Evans and Z.~Guralnik,
  Phys.\ Rev.\ D {\bf 71} (2005) 126002
  doi:10.1103/PhysRevD.71.126002
  [hep-th/0504151];
  H.~Y.~Chen, K.~Hashimoto and S.~Matsuura,
  JHEP {\bf 1002} (2010) 104
  doi:10.1007/JHEP02(2010)104
  [arXiv:0909.1296 [hep-th]]; A.~F.~Faedo, D.~Mateos, C.~Pantelidou and J.~Tarrio,
  JHEP {\bf 1710} (2017) 139
  doi:10.1007/JHEP10(2017)139
  [arXiv:1707.06989 [hep-th]].
	
\bibitem{Ramamurti:2018evz}
  A.~Ramamurti, E.~Shuryak and I.~Zahed,
  arXiv:1802.10509 [hep-ph].
	

	\bibitem{FandM}
	Freedman and McLerran, Phys. Rev. D16 (1977) 1130; Phys.
Rev. D16 (1977) 1147;
Phys. Rev. D16 (1977) 1169.
	
	
\bibitem{Faulkner:2010gj}
  T.~Faulkner, G.~T.~Horowitz and M.~M.~Roberts,
  JHEP {\bf 1104} (2011) 051
  doi:10.1007/JHEP04(2011)051
  [arXiv:1008.1581 [hep-th]].
	
	
\bibitem{evanskim}
	N.~Evans and K.~Y.~Kim,
  Phys.\ Rev.\ D {\bf 93} (2016) no.6,  066002
  doi:10.1103/PhysRevD.93.066002
  [arXiv:1601.02824 [hep-th]].
	
\bibitem{Alford:1998mk}
  M.~G.~Alford, K.~Rajagopal and F.~Wilczek,
  Nucl.\ Phys.\ B {\bf 537} (1999) 443
  doi:10.1016/S0550-3213(98)00668-3
  [hep-ph/9804403].
	
\bibitem{Witten:2001ua}
  E.~Witten,
  ``Multitrace operators, boundary conditions, and AdS / CFT correspondence,''
  hep-th/0112258.
	
\bibitem{Breitenlohner:1982jf}
  P.~Breitenlohner and D.~Z.~Freedman,
  Annals Phys.\  {\bf 144} (1982) 249.
  doi:10.1016/0003-4916(82)90116-6
	
\bibitem{Filev:2007gb}
  V.~G.~Filev, C.~V.~Johnson, R.~C.~Rashkov and K.~S.~Viswanathan,
  JHEP {\bf 0710} (2007) 019
  [hep-th/0701001].
	
	
	
	
	
	
	
	
	\end{thebibliography}
\end{document}